\documentclass[%
 reprint,
 amsmath,amssymb,
 aps,
]{revtex4-2}

\usepackage[english]{babel}

\usepackage{amsmath}
\usepackage{graphicx}
\usepackage{color}
\usepackage[dvipsnames]{xcolor}
\usepackage[colorlinks=true,
            linkcolor=black,
            citecolor=NavyBlue,
            filecolor=NavyBlue,
            urlcolor=NavyBlue
            ]{hyperref}
\usepackage{glossaries}

\setacronymstyle{long-short}
\newacronym{ar}{AR}{aspect ratio}
\newacronym{cd}{CD}{circular dichroism}
\newacronym{cpl}{CPL}{circularly polarized light}
\newacronym{fdtd}{FDTD}{finite-difference time-domain}
\newacronym{hid}{HID}{high index dielectric}
\newacronym{oce}{OCE}{optical chirality enhancement}
\newacronym{vswf}{VSWF}{vector spherical wave function}
\newacronym{tfsf}{TFSF}{total-field/scattered-field}
\makeglossaries

\begin{document}

\title{Local versus bulk circular dichroism enhancement by achiral all-dielectric nanoresonators}
\author{Krzysztof M. Czajkowski}
\email{krzysztof.czajkowski@fuw.edu.pl}
\affiliation{Faculty of Physics, University of Warsaw, Pasteura 5,PL-02-093 Warsaw, Poland}
\author{Tomasz J. Antosiewicz}
\email{tomasz.antosiewicz@fuw.edu.pl}
\affiliation{Faculty of Physics, University of Warsaw, Pasteura 5, PL-02-093 Warsaw, Poland}

\begin{abstract}
   Large optical chirality in the vicinity of achiral high index dielectric  nanostructures has been recently demonstrated as useful means of enhancing molecular circular dichroism. We theoretically study the spatial dependence of optical chirality enhancement in the vicinity of high index dielectric nanodisks and highlight its importance for the design of nanophotonic platforms for circular dichroism enhancement. Using a T-matrix framework, we demonstrate that depending on disk aspect ratio chirality is enhanced preferentially along different directions. We employ various statistical procedures (including surface, volume and orientation averaging) that are necessary to predict the enhancement of chiroptical effects and show that optimal properties of a nanostructure depend substantially on whether spatial maximum or average chirality enhancement is sought after. Similarly, the optimal choice of the nanostructure is influenced by the presence of substrate, which limits the space available to be occupied by analyte molecules and impacts the optical chirality in the vicinity of the nanostructure.
\end{abstract}
\maketitle

\section{Introduction}
Chirality is a feature of many biologically active molecules including drugs \cite{Nguyen2006}. A molecule is chiral if its mirror image is non-superimposable on the molecule by any combination of rotations and translations. Enantiomers, two configurations of a chiral molecule, may exhibit drastically different biological activity. At an extreme, one enantiomer may have beneficial medicinal effects, while the other may be toxic \cite{Smith2009}. Thus, determining the molecular helicity is of paramount important. An optical way to differentiate between enantiomers is \gls{cd} spectroscopy, which relies on differential absorption of left- and right-handed circularly polarized light. The intrinsic molecular \gls{cd} is low which leads to long acquisition times, decreased sensitivity and in turn diminishes the practicality of the method. At the same time, sensing of molecular chirality is of paramount importance e.g. in drug development. As shown by Tang and Cohen \cite{Tang2010,Tang2011}, one approach to achieve the goal of substantial \gls{oce} 
is using nanoantennas that locally enhance electromagnetic fields. This enhancement has been recently experimentally demonstrated to be useful for various types of optical spectroscopies utilizing chiroptical effects \cite{Kakkar2020,Beutel2021}.  Assuming that the molecule is a small object that can be approximated by a non-radiating point dipole, its \gls{cd} signal is proportional to the chiral part of its polarizability and optical chirality density

The figure of merit in circular dichroism enhancement is the enhancement of chirality density ($f(\vec{r})$), which is given by 
\begin{equation}
    f(\vec{r})= -Z \frac{\mathrm{Im}(\vec{E}^{*}(\vec{r}) \cdot \vec{H}(\vec{r}))}{|\vec{E}_0|^2},
\end{equation}
where $Z$ is the relative impedance of the surrounding medium, $\vec{E}$ is the electric field, $\vec{H}$ is the magnetic field and $\vec{E}_0$ is the incident electric field amplitude. We assume here that the incident field is left-handed polarized plane wave (hence the minus sign).
Recently a set of design rules, which a suitable \gls{cd}-enhancing nanostructure should fulfill, have been proposed \cite{Graf2019a}. The nanostructure should be achiral to avoid contamination of the \gls{cd} signal of the molecule by the \gls{cd} signal nanostructure itself. Otherwise, \gls{cd} signal could be generated by achiral molecules or the signal coming from the nanoparticle could be orders of magnitude larger than that of the molecule. Other rules include helicity preservation and a strong electromagnetic response.
They can be readily elucidated by representing the electromagnetic field using helicity eigenstates  $\vec{G}^{\pm}=\vec{E}\pm i Z\vec{H}$ \cite{Bialynicki-Birula2013, Fernandez-Corbaton2016}. Then, helicity change can be defined as
\begin{equation}
    \Lambda=\frac{|\vec{G}^{-}(\vec{r})|^{2}}{|\vec{G}^{+}(\vec{r})|^{2}+|\vec{G}^{-}(\vec{r})|^{2}}
\end{equation} 
and the total intensity enhancement as
\begin{equation}
    G_{enh}=\frac{|\vec{G}^{+}(\vec{r})|^{2}+|\vec{G}^{-}(\vec{r})|^{2}}{|\vec{G}^{+}_{inc}(\vec{r})|^{2}},
\end{equation} 
where $G^{+}_{inc}$ corresponds to the incident field. Using helicity eigenstates, the enhancement of optical chirality density is given by
\begin{equation}
    f =\frac{|\vec{G}^{+}|^2-|\vec{G}^{-}|^2}{|\vec{G}_{inc}^{+}|^2}=G_{enh}(1-\Lambda),
\end{equation} 
which highlights the design rules proposed by Graf et al. \cite{Graf2019a}.

Since the pioneering works by Tang and Cohen \cite{Tang2010,Tang2011} and Graf et al. \cite{Graf2019a}, several nanostructure-based concepts to enhance optical chirality density emerged. Initially, plasmonic nanostructures \cite{Luo2017} have been used, including single gold nanoparticles \cite{Besteiro2017}, gold nanoparticle dimers \cite{Zhang2013}, racemic nanoplasmonic arrays \cite{Garcia-Guirado2018}. After the publication by Graf et al. \cite{Graf2019a}, many nanostructures aiming at helicity preservation have emerged, such as Fabry-Perot cavities containing nanodisk arrays \cite{Feis2020,Beutel2021}, metalic-dielectric nanoparticle dimers \cite{Mohammadi2021}, 1D silicon nanoparticle arrays \cite{Lasa-Alonso2020}, silicon nanorods \cite{Jeon2021} and silicon nanodisks \cite{Du2021} to simultaneously attain helicity preservation and a strong electromagnetic response. A simpler route to helicity preservation is offered by isolated high-index dielectric nanoparticles such as silicon nanospheres \cite{Ho2017,Garcia-Etxarri2013,Garcia-Guirado2020} or nanodisks \cite{Solomon2019}. In the dipolar approximation minimization of helicity change requires fulfilling the Kerker conditions, such as equal electric and magnetic dipole scattering amplitudes \cite{Zambrana-Puyalto2013}. Consequently, \gls{hid} nanodisks with overlapping electric and magnetic resonances have been considered as optimal for optical chirality density enhancement.

At the same time, one has to note that the optical chirality density is a spatially dependent property, but \gls{cd} is a macroscopic property determined by the integral of chirality density over the volume outside the particle
\begin{equation}
    CD = \frac{\text{Im}(\alpha_{em}) \omega}{2 c} \int_{V} f(\vec{r})  dV',
\end{equation}
where $V$ is the volume occupied by the molecules.
Chirality density averaged over a spherical surface has been proposed by Garcia-Etxarri and Dionne \cite{Garcia-Etxarri2013} as a close proxy of the bulk enhancement
\begin{equation}
    f_{avg}(r) = \frac{1}{4\pi} \int_0^{2\pi} d\phi \int_0^{\pi} f(\vec{r}) \sin \theta  d\theta.
    \label{eq:favg}
 \end{equation}
Due to spatial dependence of $f(\vec{r})$, the attainable $CD$ enhancement depends on the spatial distribution of the molecules and thus designing the nanostructure for $CD$ enhancement might depend on the specific experimental scenario. 

Here, we combine semi-analytical considerations based on T-matrix formalism and numerical study using \gls{fdtd} method to find out effective means of enhancing electromagnetic chirality using an archetypical all-dielectric nanoresonator -- namely, a dielectric nanodisk. We highlight and focus on a considerable difference between local and bulk/surface averaged enhancement of chirality and study the dependence of the optical chirality density enhancement for a nanodisk geometry for each type of spatial averaging or lack of such averaging. We show that the optimal nanodisk aspect ratio depends on the spatial extent in which the chiral molecules reside and that at times sacrificing helicity preservation for a stronger electromagnetic response in the volume occupied by the molecules is beneficial for optical chirality enhancement.

This work is structured as follows. First, we describe the theoretical approach used herein. Afterwards, we show that the direction along which chirality is enhanced and the maximal attainable chirality enhancement along each direction depends on the particle shape. We establish a simple semianalitical formalism based on the dipole approximation which is able to rationalize the result. Next, we use the T-matrix method to find the surface averaged and orientation averaged chirality enhancement and combine these results with helicity-preserving multipolar decomposition to find out the dependence of chirality enhancement on particle aspect ratio and helicity change upon scattering. Finally, we use the \gls{fdtd} method to find local chirality enhancement maxima as well as averages of chirality enhancement over thin layers placed around the particle. We also study the impact of placing the particle on a substrate on anticipated chirality enhancement, because the presence of a substrate is a common factor in experimental platforms for nanophotonic enhancement of molecular circular dichroism.

\section{Theory}
As mentioned in the introduction optical chirality density enhancement $f$ is a useful figure of merit in circular dichroism. Here, we outline the theoretical framework enabling calculating chirality density using the scattering problem perspective and helicity preserving multipoles defined \cite{Graf2019a}. Both $\vec{G}^{+}$ and $\vec{G}^{-}$ are separated into incident and scattered fields
\begin{equation}
    \vec{G}^{\pm}=\vec{G}_{inc}^{\pm}+\vec{G}_{scat}^{\pm}.
\end{equation}
Then the optical chirality enhancement can be split into interference ($f_{int}$) and scattered parts ($f_{scat}$), \cite{Hanifeh2020a}
\begin{equation}
    f=1+f_{int}+f_{scat}
    \label{eq:enhscatprob}
\end{equation}
with 
\begin{equation}
    f_{int}= \frac{2 \text{Re} (\vec{G}_{scat}^{+} \cdot \vec{G}_{inc}^{+} - \vec{G}_{scat}^{-} \cdot \vec{G}_{inc}^{-})}{|\vec{G}_{inc}^{+}|^2}
\end{equation}
and
\begin{equation}
    f_{scat}=\frac{1/4(|\vec{G}_{scat}^{+}|^2-|\vec{G}_{scat}^{-}|^2|)}{|\vec{G}_{inc}^{+}|^2}
\end{equation}
Note that $f_{scat}$ is strictly positive unless the structure scatters fields with right-handed helicity $(-)$ more efficiently than those with left-handed $(+)$ helicity. To the contrary, $f_{int}$ has a plus or minus sign depending on the relative phase between the incident and scattered fields. 

The scattered field is found either using the T-matrix approach or dipole approximation. The T-matrix has been recently shown as a valuable tool for analysis of chiral light-matter interactions \cite{Mun2019}. In this work, open source code \texttt{SMUTHI} is used to easily calculate the T-matrix and related quantities \cite{EGEL2021107846}. The T-matrix framework facilitates finding field distributions outside the circumscribing sphere of the particle by relying on field expansion using \glspl{vswf} \cite{Doicu2006}. Due to the fact that we consider here chiral properties of optical fields, we use multipoles of pure helicity
\begin{equation}
    b^\pm_{l,m} =\frac{b^e_{l,m}\pm b^m_{l,m}}{\sqrt{2}}.
\end{equation}
Using these multipole moments it is possible to calculate helicity change upon scattering simply as \cite{Graf2019a}
\begin{equation}
\Lambda = \frac{\sum_{l,m} |b^-_{l,m}|^2}{\sum_{l,m} |b^-_{l,m}|^2 + |b^+_{l,m}|^2}.  
\label{eq:helicity}
\end{equation}
Another advantage of using \glspl{vswf} for field expansion is their orthogonality on a spherical surface \cite{Doicu2006} and hence the integral in \autoref{eq:favg} defining the surface averaged chirality enhancement can be evaluated analytically. We assume that the incident field is a left handed circularly polarized plane wave with its wavevector aligned with the z-axis. We decompose $f_T$ using \autoref{eq:enhscatprob} and find the interference ($f_{int}^{T}$) and scattering ($f_{scat}^{T}$) parts separately. In this case, the interference part is
\begin{equation}
    f_{int}^{T}=\frac{1}{2} \sum_{l} \text{Im}\left(i^{l-1} w_{ext,l}(k,r) b_l^{+}\right)
\end{equation}
with 
\begin{multline}
    w_{ext,l}(k,r)=j(kr) h(kr)(1+\frac{l(l+1)}{k^2r^2})+ \\ \frac{1}{k^2r^2}\frac{\text{d}}{\text{d}kr} [kr h_l(kr)] \cdot \frac{\text{d}}{\text{d}kr} [kr j_l(kr)].
\end{multline}
The scattering part is 
\begin{equation}
    f_{scat}^{T}=\frac{1}{8} \sum_{l} w_{scat,l}(k,r)\left(|b_l^+|^2-|b_l^{-}|^2\right)
\end{equation}
with 
\begin{multline}
    w_{scat,l}=|h(kr)|^2(1+\frac{l(l+1)}{k^2r^2})+ \\ \frac{1}{k^2r^2}\left|\frac{\text{d}}{\text{d}kr} [kr h_l(kr)]\right|^2.
\end{multline}

This result can be then further averaged over nanoparticle orientation using the concept of the orientation-averaged T-matrix ($\langle T \rangle$) defined by Doicu et al \cite{Doicu2006}. Average $f_{int}$ is easily evaluated as
\begin{equation}
   \langle f_{int}^{T} \rangle=2 \sum_{l} \text{Im}(i^{l-1} w_{ext,l}(k,r) \langle b_l^{+} \rangle )
    \label{eq:orientationavext}
\end{equation}
with $\langle \vec{b} \rangle$ defined as $\langle \vec{b} \rangle = \langle T \rangle \vec{a}$, where $\vec{a}$ is the incident field coefficient vector.
The scattered part is defined as 
\begin{equation}
   \langle f^T_{scat} \rangle = \vec{a}^{\dagger} \langle W \rangle \vec{a}.
    \label{eq:orientationavscat}
\end{equation}
Matrix $W$ is defined as $W=T^{\dagger} F T$ with $    F = \begin{bmatrix}
0  & |h(kr)|^2 \\
\frac{1}{k^2r^2}\left|\frac{\text{d}}{\text{d}kr} [kr h_l(kr)]\right|^2+l(l+1)|\frac{h(kr)}{kr}|^2 &    0
    \end{bmatrix}$. Its orientation average is given by
\begin{equation}
    \langle (W)^{i,j}_{l,m,l',m'} \rangle =\delta_{m m'}\delta_{l l'}\hat{t}^{ij}_l
\end{equation}
with
\begin{equation}
    \hat{t}^{ij}_l=\frac{1}{2l+1}\sum_{m'} (W)^{ij}_{l,m,l',m'}
\end{equation}
Here, $i$ and $j$ index the dipole types ($mag$ and $el$). The details of orientation averaging procedure for $f^T$ are given in Appendix B.

As shown in the previous paragraphs, the spherical basis is useful for expressing surface and orientation averages of optical chirality enhancement. Directional \glspl{oce} are more readily evaluated in the Cartesian basis. Here, we resort to the dipole approximation as it allows for relatively compact expressions that describe the \gls{oce} along specified directions and its value when averaged over a spherical surface. Instead of the T-matrix, polarizability is used in the dipolar approximation to describe the nanoparticle properties. The relationship between electric/magnetic polarizability (in CGS units) and the corresponding T-matrix element is $\alpha_{e/m} = -i \frac{3}{2 k^3}T_{ee/mm}$. Polarizability is transformed to multipoles with defined helicity where $\alpha^{\pm}=\frac{\alpha^e \pm \alpha^m}{\sqrt{2}}$. The enhancement along $x$ and $y$ directions can be calculated as 
\begin{multline}
   f_{xy}= 1 + \left(|\alpha^{+}|^2-|\alpha^{-}|^2\right)w_{xy}+ \\ \text{Re}\left(\alpha^{+}\frac{ \exp(i k r) \left(k^2 r^2+1 -ik r\right)}{\sqrt{2} r^3}\right),
   \label{eq:fxy}
\end{multline}
with $w_{xy}=\frac{2 k^4 r^4+4 k^2 r^2+5}{4 r^6}$.
The enhancement along $z$ is
\begin{multline}
 f_{z}= 1+ \left(|\alpha^{+}|^2-|\alpha^{-}|^2\right)\frac{1}{2 r^6}+|\alpha^{+}|^2\frac{ 4 k^4 r^4}{2 r^6}+ \\ \text{Re}\left(\alpha^{+} \frac{\sqrt{2}  \left(2 k^2 r^2-1\right)+i2 \sqrt{2}  k r}{r^3} \right).
 \label{eq:fz}
\end{multline}
The dipolar approximation can be also used to perform surface averaging, giving
\begin{multline}
f_{avg}= 1+ w_{avg}(k,r) \left(\left(|\alpha^{+}|^2-|\alpha^{-}|^2\right) - \frac{3\text{Im}(\alpha^{+})}{\sqrt{2}k^3}\right) + \\ \text{Re}\left(\alpha^{+} \frac{\exp{(2ikr)} \left(-6 k r + i \left(3-4 k^2 r^2\right)\right) }{\sqrt{2} k^3 r^6}\right) 
\label{eq:favg}
\end{multline}
with $w_{avg}(k,r) = \left(2 k^4 r^4+2 k^2 r^2+3\right)/(3 r^6)$.
Note, that this extends other dipolar approximations present in the literature \cite{Lee2017,Garcia-Etxarri2013,Hanifeh2020} by not resorting to the quasi-static approximation and considering \gls{oce} along specific directions and its surface average.


\section{Results and discussion}
\begin{figure}
    \centering
    \includegraphics[width=\columnwidth]{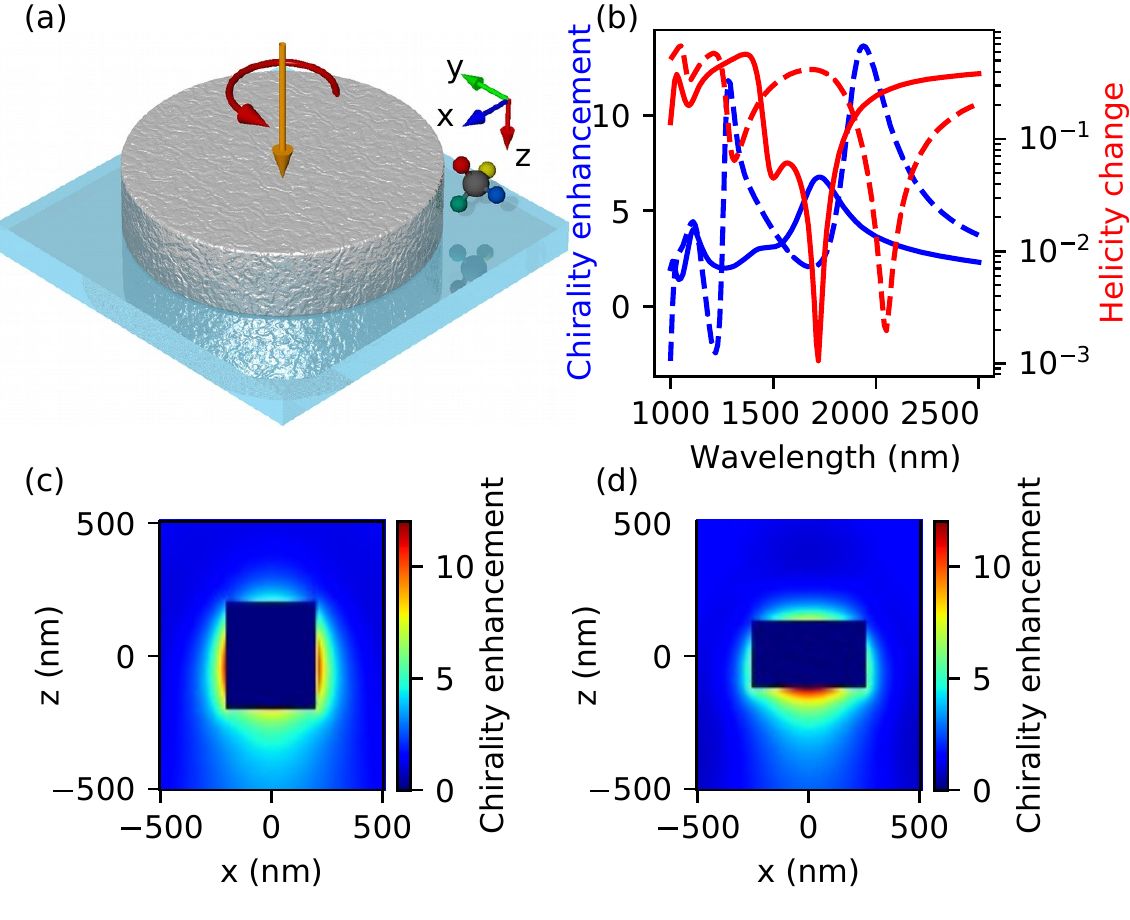}    
    \caption{ (a) Schematic representation of the studied system. High index ($n=4$ to mimic silicon) dielectric nanodisk is used to enhance the optical chirality density at molecule location. The system is illuminated by a left-handed circularly polarized light and embedded in water ($n=1.33$). (b) Comparison of helicity change and chirality enhancement spectra for the two disks with varying AR. Solid lines -- $\mathcal{A}=0.5$, dashed lines -- $\mathcal{A}=1$. Chirality enhancement is averaged over a spherical surface with radius 50~nm larger than that of a disk. For the disk with $\mathcal{A}=0.5$ maximal enhancement is matched with helicity change minimum, while for $\mathcal{A}=1$ the two are significantly shifted with respect to each other. Maximum surface averaged enhancement for the disk with $\mathcal{A}=1$ is larger than for the disk with $\mathcal{A}=0.5$, yet it corresponds to large helicity change on the order of 10\%. Spatial maps of chirality enhancements for nanodisks with ARs (c) $\mathcal{A}=1$ and (d) $\mathcal{A}=0.5$. Note how, $\mathcal{A}=0.5$ leads to enhancement predominantly along the $z$-axis below the particle, while when $\mathcal{A}=1$ chirality is enhanced preferentially around the middle (circumference).  }
    \label{fig:maps}
\end{figure}

Dielectric nanodisks provide local enhancement of electromagnetic chirality, which in turn results in enhancement of circular dichroism of molecules localized in the vicinity of such nanodisks. \autoref{fig:maps} presents maps of local enhancement of chirality for two exemplary \gls{hid} nanodisks with equal volume of $V=5 \times 10^7$ nm$^3$ and different \gls{ar} defined as $\mathcal{A}=H/D$ with $H$ being the thickness and $D$ the diameter of the nanodisk. The illumination is plane wave propagating in the $-z$-direction with right handed circular polarization. Its wavelength is tuned for each disk to obtain maximal local chirality enhancement. The maximum attainable local enhancement of up to 12 is provided by a nanodisk with \gls{ar} of about 0.5. The optical response of this nanodisk consists of overlapping electric and magnetic dipole resonances, which fulfills the Kerker condition, while maintaining a strong resonant electromagnetic response. Such disks have been considered optimal for chirality enhancement. At the same time, we observe that when \gls{ar} is changed so that the disk thickness is equal to its diameter, the maximal local enhancement does not change substantially (see \autoref{fig:maps}b). Instead, other changes are more noticeable. Namely, the locations at which chirality enhancement is maximized are very different for each of the two analyzed cases. For disk with $\mathcal{A}=0.5$, the maximum is located below the particle in the $z$-direction, which coincides with the propagation direction of the incident wave. In contrast, for disk with $\mathcal{A}=1$, the maximum is located at the sides of the disk around its circumference. Enhancement at the top and bottom surfaces of this disk is substantially lower (around 7). 

To quantify chirality enhancement without having to specify the location, we use $f_{avg}$. The average chirality enhancement spectra for both nanodisks under consideration are presented in \autoref{fig:maps}c along with corresponding helicity change spectra calculated using the T-matrix method. The size of the spherical shell is set so that the distance between particle surface and the shell surface is 50 nm. Despite the fact that the maximal attainable local enhancement is provided by nanodisk with AR~=~0.5, it is the other nanodisk that enables higher average chirality enhancement maximum. 

\begin{figure}
    \centering
    \includegraphics[width=\columnwidth]{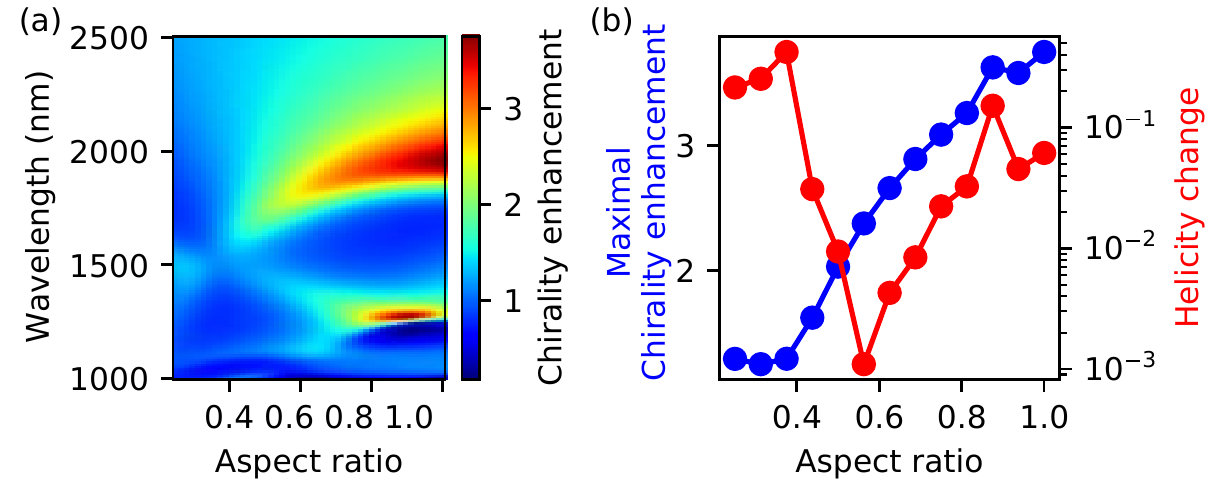}    
    \caption{(a) Surface averaged chirality enhancement spectra as a function of aspect ratio $\mathcal{A}$ for water-embedded HID nanodisks with fixed volume of 5$\times 10^7$ nm$^3$. For flat disks the enhancement is generally low. For tall disks two contributions are present: dipolar around 1750-2000 nm and quadrupolar around 1250 nm. (b) Maximal surface averaged chirality enhancement as a function of AR and corresponding helicity change. Note how chirality enhancement is monotonic with aspect ratio (except near AR=1.75, where quadrupole mode leads to maximal enhancement), while it is not correlated with the corresponding helicity change.}
    \label{fig:average}
\end{figure}

Next, we investigate how chirality enhancement and helicity change based on the far-field calculation vary over a broad range of nanodisk aspect ratio. To that end we utilize the T-matrix formalism which facilitates rapid calculation of chiral properties of the electromagnetic field scattered by nanoresonators as shown in the Theory section. 
The surface averaged chirality enhancement spectrum as a function of nanodisk aspect ratio is presented in \autoref{fig:average}a. For flat disks with \gls{ar} below 0.4 the observed enhancement is almost negligible. Above this value we observe that with increasing \gls{ar} the surface averaged chirality enhancement increases. Furthermore, for nanodisks with sufficiently large \gls{ar} two maxima in the chirality enhancement spectrum are present. The one corresponding to the larger wavelength results from an interplay of electric and magnetic dipoles, while the second one corresponds to quadrupoles. The quadrupolar peak is narrower than the dipolar one, which is typical for optical cross-section spectra of \gls{hid} nanodisks. Depending on \gls{ar}, the quadrupolar peak might have a larger amplitude than the dipolar one, but it is worth noting that the considered structures lack absorption (having used $n=4$ only), which tends to diminish the impact of quadrupoles on the optical properties (although for Si this would happen only for significantly shorter wavelengths).

In \autoref{fig:average}b we compare the maximal chirality enhancement (maximum is taken over wavelength, while averaging is performed over the particle surface) with the helicity change calculated based on helicity preserving multipole decomposition (see \autoref{eq:helicity}) at the wavelength corresponding to the chirality enhancement maximum. The helicity change features a very clear minimum at \gls{ar} of approximately 0.55, where its value is about 0.1 \%. This due to the fact that the disk with this \gls{ar} fulfills the Kerker condition, which is known to be the condition for minimal helicity change \cite{Zambrana-Puyalto2013}. An increasing value of helicity change is observed as \gls{ar} shifts away from the value corresponding to the Kerker condition. Over 10 \% of helicity change is observed for very small aspect ratios and for \gls{ar} of ca. 0.9. While no clear correlation between surface averaged chirality enhancement and helicity change is observed, elongated cylinders perform better than flat disks and not even a local maximum is observed for the \gls{ar} corresponding to the helicity change minimum. Surface averality enhannt almost exclusively increases with increasing \gls{ar}. The exception is a local maximum around \gls{ar} of 0.9, which stems from a narrow quadrupole peak. This peak gives the maximum chirality enhancement while simultaneously yielding an exceptionally high helicity change.

\begin{figure}
    \centering
    \includegraphics[width=\columnwidth]{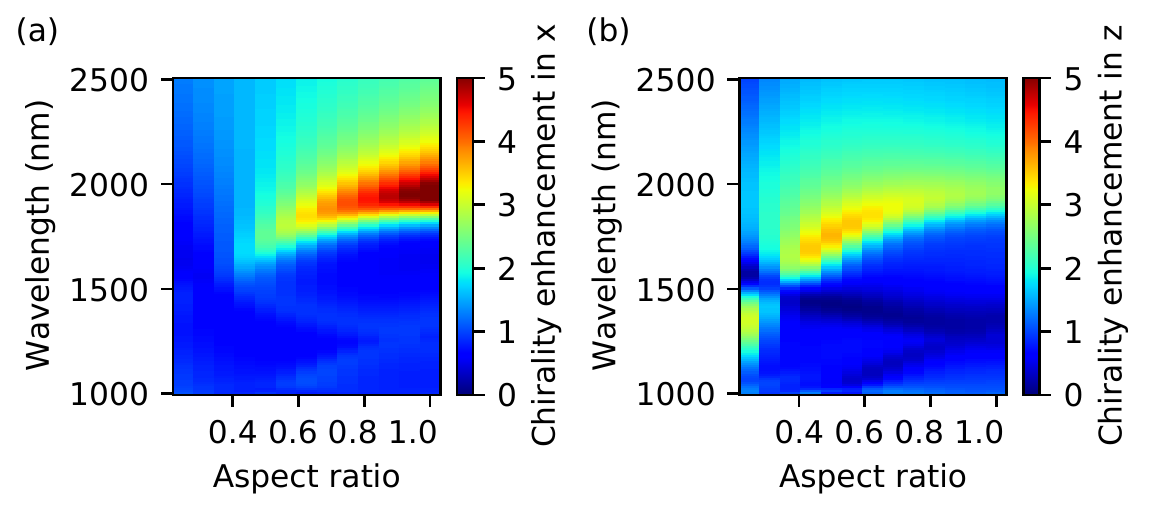}    
    \caption{Chirality enhancement spectra for high-index nanodisks calculated along specific directions: (a) in $x$, (b) in $z$ at 50 nm away from each particle. Here, semi-analytical expressions obtained in the dipole approximation are used. Therefore, quadrupole maxima are not present. Note how maxima of chirality enhancements appear for different aspect ratio for each direction. Also, maximal attainable value for the $z$ direction is smaller than that in the $x$ direction.}
    \label{fig:directionality}
\end{figure}

While the proposed T-matrix method based formalism enables fast calculation of the surface averaged chirality and helicity change, it does not provide simple analytical formulas that bring additional physical insight. To that end, we resort to dipolar approximation, which enables calculation of optical chirality enhancement along the $x$- and $z$-axis via \autoref{eq:fxy} and \autoref{eq:fz}, respectively. \autoref{fig:directionality} presents the dependence of the chirality enhancement spectrum on the nanodisk aspect ratio for $x$ and $z$ directions. We fix the distance between the nanodisk surface and the observation point (molecule) along $x$ at 50~nm. The result obtained for \gls{oce} calculated along $x$ is qualitatively similar to the surface averaged chirality enhancement obtained using the T-matrix approach (shown in \autoref{fig:average}), with the exception of the quadrupolar feature at 1300 nm, which is omitted in this qualitative description. 

The results obtained for points along $x$ and $z$ directions are qualitatively very different (cf. \autoref{fig:directionality}ab). The chirality enhancement along the $x$ direction is maximized by nanodisks with large aspect ratios following a similar dependence to that for the surface averaged chirality. In contrast, the enhancement along $z$ is maximized by choosing an aspect ratio around 0.5, which corresponds to fulfilling the Kerker condition. With increasing aspect ratio, the chirality enhancement decreases while maintaining a similar wavelength dependence to those observed for surface averaged chirality. These observations are in agreement with the \gls{fdtd} calculations presented in \autoref{fig:maps}. Both methods lead to the conclusion that for small aspect ratios the enhancement is predominantly in $z$, while for larger aspect ratios it is observed predominantly in the $x$ direction. The enhancements observed along $z$ are smaller than that in $x$ in dipolar calculations, while the converse is true for \gls{fdtd} calculations. We attribute this difference to the fact that when the dipole approximation is used the fields along $z$ are calculated above the particle, while the enhancement is more prominent below the particle.

\begin{figure*}
    \centering
    \includegraphics{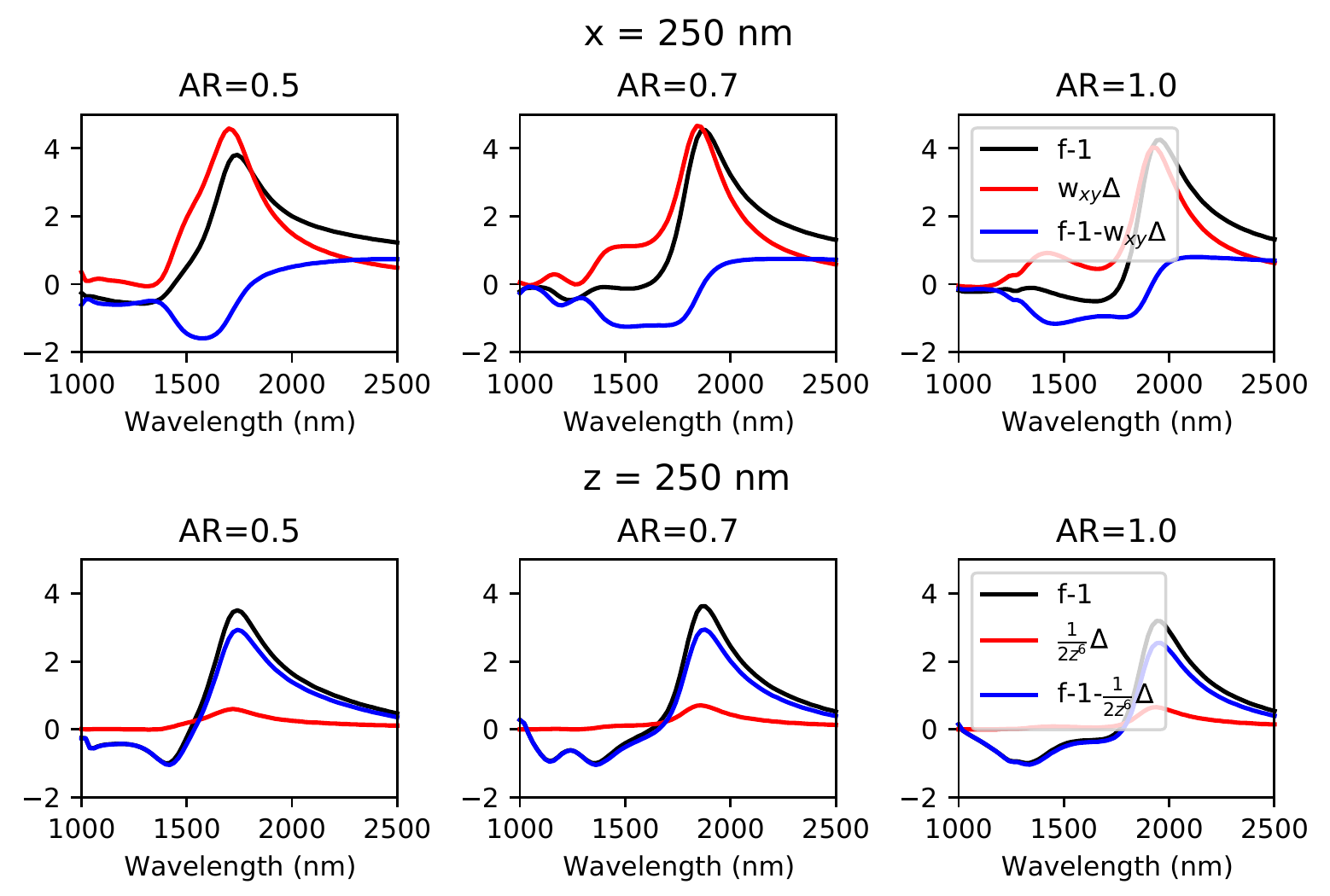}
    \caption{Various contributions to chirality enhancement ($f$) spectra in dipolar approximation: chirality enhancement minus one, term proportional to $\Delta=|\alpha^{+}|^2-|\alpha^{-}|^2$ and the remaining part that depends on $\alpha^{+}$ (independent on $\alpha^{-}$). Selected values of aspect ratio are being used. Top: at a single point with $x=250$~nm, at a single point with $z=250$~nm.}
    \label{fig:dipoleapprox_newfig}
\end{figure*}

As readily seen from \autoref{eq:fxy} and \autoref{eq:fz}, placing the sensed entity along different axes has a significant impact on the individual terms to optical chirality enhancement, which depends on the wavenumber and nanoparticle-molecule center-to-center distance. Here, we fix the displacement of the molecule along $x$ or $z$ at 250~nm and present the wavelength dependence of \gls{oce} on two relevant terms: the one proportional to $\Delta(\equiv|\alpha^{+}|^2-|\alpha^{-}|^2)$ and the reminder term. These data are plotted in \autoref{fig:dipoleapprox_newfig} for selected aspect ratios of the nanodisk. 
Observation of large \gls{oce} requires simultaneously a large $\Delta$-proportional term and interference (reminder) term. 
As seen in the top panel of the figure, this is not true for the Kerker disk, which offers the largest $\Delta$ proportional term and a large negative interference term, which diminishes substantially the observed enhancement. With the position of the molecule being fixed (instead of the particle surface-molecule distance), the \gls{oce} is almost independent of the aspect ratio as with increasing aspect ratio the $\Delta$ proportional term decreases, but the interference term increases to compensate. Eventually, for aspect ratio of 1, the interference term is positive.

The decomposition of optical chirality enhancement along $z$ reveals very different trends than in the $x$-axis case. The $\Delta$-proportional term decays with the center-to-center distance as $\frac{1}{r^6}$, which means that it contributes very little to the actual result. This is clearly seen in the bottom panel of \autoref{fig:dipoleapprox_newfig}. This is a rather striking observation. It indicates that $\alpha^{+}$ rather than $\Delta$ determines the magnitude of \gls{oce}. Consequently, \gls{oce} does not require as strict a helicity preservation as it would be if the $\Delta$-proportional term would be dominant. It also means that the overlap of electric and magnetic resonances is useful for obtaining large $\alpha^{+}$ rather than producing a large contrast between $\alpha^+$ and $\alpha^-$ in this case. It can be equally well replaced by a particle that simply features a strong $\alpha^+$ response. As an example, plasmonic particles that do not feature a magnetic dipole response would have $\alpha^{+}=\alpha^{-}=\alpha_e$, which would lead to equal amounts of $+$ and $-$ helicity being produced by scattering of light with $+$ helicity. Hence, $\Delta$ would equal to zero. At the same time, $\alpha^{+}$ can be large and lead to substantial optical chirality enhancement.

\begin{figure}
    \centering
    \includegraphics{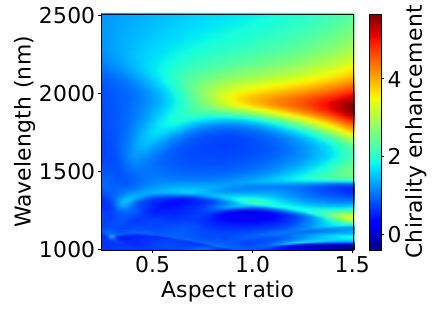}
    \caption{Orientation averaged chirality enhancement as a function of $\mathcal{A}$ obtained using the T-matrix method. In this scenario, chirality is enhanced predominantly by dipole modes with maximal enhancements observed for particles with large aspect ratios.}
    \label{fig:orientation}
\end{figure}

The strength of the T-matrix method lies also in its ability to yield optical quantities averaged over nanoparticle orientations \cite{Fazel-Najafabadi2021}. Orientation averaging is especially useful for predicting the properties of nanoparticles in solutions, in which nanoparticles are of arbitrary orientation and thus orientation averaging is necessary for this task. Here, we utilize the tools developed in the Theory section to (see \autoref{eq:orientationavext} and \autoref{eq:orientationavscat}) to find the dependence of orientation averaged chirality enhancement on the nanodisk aspect ratio. The results of this study are presented in \autoref{fig:orientation}. The size of the spherical surface over which spatial averaging is performed is determined such that the distance between the nanodisk surface and the sphere along $x$ is 50 nm. We observe that the orientation averaged \gls{oce} follows similar trends as surface averaged chirality presented in \autoref{fig:average}. At the same time, while surface averaged chirality featured a prominent quadrupolar feature for large aspect ratio, this feature is much less intense when averaged over particle orientation. 

The main limitation of the T-matrix method (also in the dipolar approximation) is that it does not 
yield reliable values of the 
near fields located within the smallest circumscribing sphere of a nanoparticle. Thus, it is not possible to perform a volume integral to calculate volumetric circular dichroism enhancement. Also, the method of surface averaging proposed here does not extend easily to particles on a substrate as it would require accounting for substrate-mediated self-coupling. To circumvent these issues we complement the T-matrix method by  \gls{fdtd} simulations, which are substantially more resource- and time-consuming, but are reliable enough to yield accurate near fields and particles placed on a substrate. 

\begin{figure}
    \centering
    \includegraphics[width=\columnwidth]{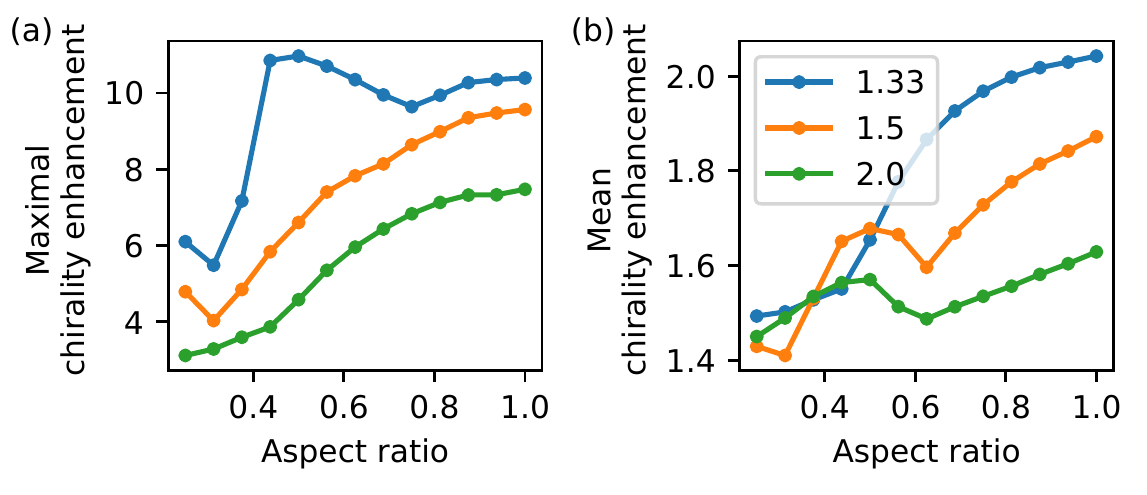}
    \caption{(a) Maximal and (b) mean chirality enhancement as a function of AR for varying substrate refractive index: index-matched (1.33), glass (1.5) and high-index dielectric (2). Here, the statistics are taken over the upper half-space of the simulation, with the exception of index-matched case in which the both half-spaces are used (assuming that in a solution the specimen can take arbitrary position, while in the presence of a substrate, molecules cannot enter the substrate). The space occupied by the particle is always excluded from the analysis.}
    \label{fig:fdtdmaximalmean}
\end{figure}

We use a total-field/scattered-field source and monitor both the electric and magnetic fields inside a box with a side length of 500~nm, which is large enough to fit all disks with various \glspl{ar} without changing its size. As a figure of merit of bulk \gls{cd} enhancement we use the chirality enhancement averaged over the box volume excluding the particle volume and the substrate. In a homogeneous medium the integration extends also over the bottom half-space. We compare such a volumetric mean of \gls{oce} to its spatial maximum over the integration volume in \autoref{fig:fdtdmaximalmean}. Clearly the maximal attainable local enhancement varies greatly from the volume average. This result confirms the validity of the T-matrix method based analysis presented above. Local \gls{oce} in a homogeneous environment is maximized by a Kerker disk as shown by other authors earlier \cite{Solomon2019,Graf2019a}. This local maximum is found below the particle as indicated in \autoref{fig:maps}a. Nanodisks with larger aspect ratios perform worse than a corresponding Kerker disk in terms of the maximal local enhancement. However, the difference is not large -- it is around 11 for the Kerker disk and over 10 for the disk with its aspect ratio equal to 1. The situation changes when the nanodisk is placed on a substrate. Then, we assume that the molecules cannot reside in the bottom half-space, where the maximum for the Kerker disk is localized. In this case, utilizing enhancement along x-axis is more beneficial, which shows nanodisks with large aspect ratios to be better at enhancing chirality than the Kerker disk. We observe that the presence of a substrate is detrimental to chirality enhancement with maximal attainable enhancement (for nanodisk aspect ratio of 1) of about 9 for a glass ($n_{sub}=1.5$) substrate and about 7 for an index ($n_{sub}=2$) substrate.

The volume averaging results obtained with \gls{fdtd} are presented in \autoref{fig:fdtdmaximalmean}b. When the nanodisk is embedded in a homogeneous environment, the T-matrix results are qualitatively reproduced. The volume averaged \gls{oce} increases with increasing nanodisk aspect ratio. However, it reaches only a very modest value of about 2 for the largest \gls{ar} -- a value far small than of the local maximum. This indicates that nanostructures are most promising for enhancing chirality in their close vicinity, but not for enhancing the chiroptical response in bulk. The enhancement is additionally limited if the substrate is present. Especially, in this case the dependence on aspect ratio is no longer monotonic and exhibits a minimum for \gls{ar} of approximately 0.6.

\begin{figure*}
    \centering
    \includegraphics{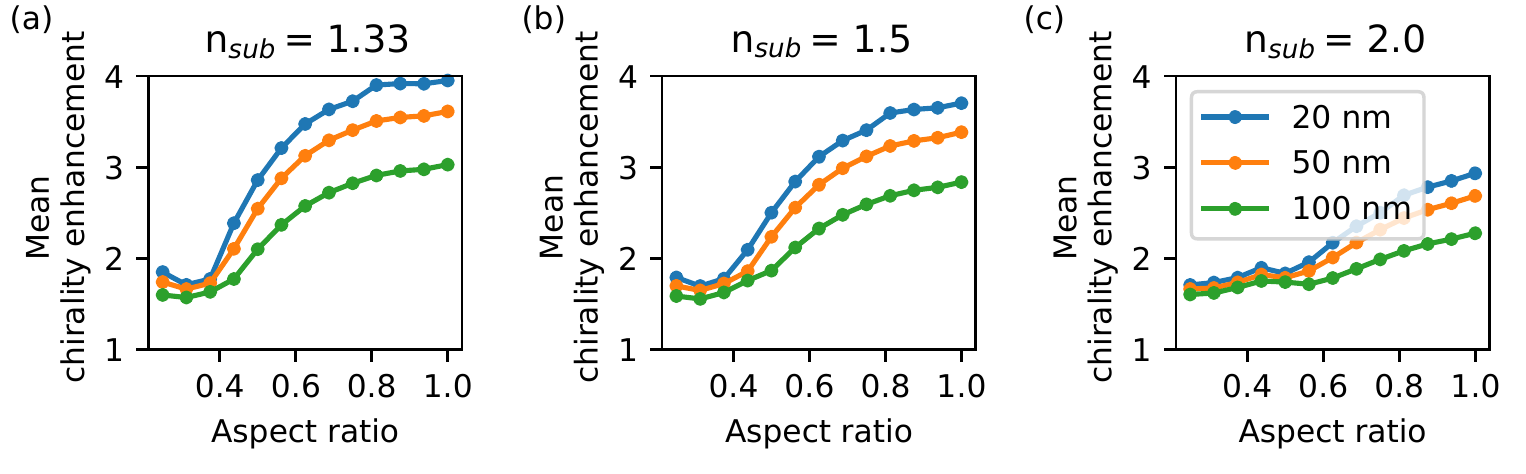}
    \caption{Mean chirality enhancement in thin layers (thickness from 20 to 100 nm) surrounding the nanodisk. (a) disk embedded in water $n_{sub}=n_{env}$, (b) glass with $n=1.5$, and (c) high-index dielectric $n=2$. It is clear that a substrate is detrimental to the device performance, while the dependence on AR is always close to that obtained by surface averaging.}
    \label{fig:fdtdlayers}
\end{figure*}

As a last step of this study, we analyze enhancement of optical chirality in a thin layer surrounding the nanodisk to mimic a monolayer adsorption experiment. 
This case resembles the one of surface averaged chirality, but here we adopt the shape of the nanoparticle when averaging. The results are presented in \autoref{fig:fdtdlayers}. We observe that the layer averaged values follow the volumetric enhancement trends presented in \autoref{fig:fdtdmaximalmean}. Because  optical chirality decays with an increasing distance from the nanodisk, thin layers result in larger mean enhancement compared to volumetric enhancement. Also, \gls{oce} decreases with increasing layer thickness. As in the other cases, the presence of the substrate limits the attainable enhancement.

\section{Conclusions}
In summary, we show that due to electromagnetic chirality being a local property with considerable spatial variation, there is a significant difference in the observed circular dichroism depending on the concentration and distribution of the analyte molecules. In terms of modelling, this implies that different statistical procedures are necessary to predict an experiment's outcome depending on how it is designed. We have compared three possible scenarios: local enhancement, surface/layer enhancement and volumetric enhancement of electromagnetic chirality using achiral high-index dielectric nanocylinders. While such nanoresonators have been widely investigated in terms of their local capabilities to enhance chirality, a more comprehensive study was lacking. Here, we demonstrate that if average enhancement is sought, the optimal properties of a nanostructure to enhance chirality are considerably different that those optimizing the spatial chirality maximum. Even when local chirality enhancement is considered, the optimal nanostructure geometry depends on the point in space in which one wants to enhance chirality. We anticipate that this work provides a useful theoretical framework for studying spatial dependence of optical chirality and that it facilitates designing of future experimental realizations of nanophotonic platforms for chiral molecule sensing.
\section*{Acknowledgements}
The authors acknowledge support by the Polish National Science Center via the project 2020/37/N/ST3/03334. Data and code supporting the findings of this study
are available at  \href{https://zenodo.org/record/6565968#.YovbKTnP02o}{doi:10.5281/zenodo.6565968}
\bibliographystyle{unsrt}
\bibliography{sample}
\end{document}


\maketitle

\section*{Appendix A: Chirality enhancement in dipolar approximation}

To calculate the chirality enhancement for scatterers in a homogeneous environment in the dipolar approximation the field is decomposed into incident and scattered components
\begin{align}
    \vec{E} &= \vec{E}_{inc}+\vec{E}_{scat}, \\
    \vec{H} &= \vec{H}_{inc}+\vec{H}_{scat}.
\end{align}
The scattered field is calculated using the Green tensor \cite{Evlyukhin2010}. We utilize CGS units for this Appendix. 
\begin{align}
    \vec{E}_{scat} &= G\vec{p}-(\vec{g}\times\vec{m}),\\
    \vec{H}_{scat} &= G\vec{m}+(\vec{g}\times\vec{p}),
\end{align}
with
\begin{equation}
    G = \frac{e^{ikr}}{r}\left(\left(k^2+\frac{ik}{r}-\frac{1}{r^2}\right)\hat{I} + \left(-k^2-\frac{3ik}{r}+\frac{3}{r^2}\right)\hat{r} \hat{r}\right)
\end{equation}
and
\begin{equation}
    \vec{g} = \frac{i e^{i k r} k (i k r-1)}{r^2}\hat{r}.
\end{equation}
Dipole moments ($\vec{p}$ and $\vec{m}$) of an achiral structure are related to the incident fields via polarizability tensors ($\alpha^e$ and $\alpha^m$).
Assuming that the field propagates in the $z$-direction and that the nanostructure enhancing chirality is axially symmetric and achiral,
\begin{align}
    \vec{p} &= \alpha^{el}_{xx} \vec{E}_{inc}, \\
    \vec{m} &= \alpha^{mag}_{xx} \vec{H}_{inc}.
\end{align}

The optical chirality density enhancement is found by evaluating $G$ and $\vec{g}$ along $x$ direction (with spherical coordinates $\theta=\frac{\pi}{2},\phi=0$) and along $z$ direction (with spherical coordinates $\theta=\phi=0$). Then, the resulting $\vec{E}$ and $\vec{H}$ are inserted in eq. 1 in the main text. In order to convert the results to helicity preserving dipoles, we utilize the relation presented in the main text,
\begin{equation}
    \alpha^\pm = \frac{\alpha^{el} \pm \alpha^{mag}}{\sqrt{2}}.
\end{equation}

\section*{Appendix B: Bulk chiral sensing in T-matrix formalism}
\subsection*{T-matrix formalism}
In the T-matrix formalism scattered electric and magnetic fields are expanded into radiating VSWFs as
\begin{equation}
  \vec{E}_{scat}(\vec{r}) = \sum_{l,m}  {b}^{mag}_{l,m} \vec{M}^{3}_{l,m}(\vec{r}) + {b}^{el}_{l,m} \vec{N}^{3}_{l,m}(\vec{r})
  \label{eq:escat}
\end{equation}
and
\begin{equation}
  \vec{H}_{scat}(\vec{r}) = \frac{1}{i Z} \sum_{l,m}  {b}^{el}_{l,m} \vec{M}^{1,3}_{l,m}(\vec{r}) + {b}^{mag}_{l,m} \vec{N}^{1,3}_{l,m}(\vec{r}),
\end{equation}
while incident electric and magnetic fields are expanded into regular VSWFs
\begin{equation}
  \vec{E}_{inc}(\vec{r}) = \sum_{l,m}  {a}^{mag}_{l,m} \vec{M}^{1}_{l,m}(\vec{r}) + {a}^{el}_{l,m} \vec{N}^{1}_{l,m}(\vec{r})
\end{equation}
and
\begin{equation}
  \vec{H}_{inc}(\vec{r}) = \frac{1}{i Z} \sum_{l,m}  {a}^{el}_{l,m} \vec{M}^{1}_{l,m}(\vec{r}) + {a}^{mag}_{l,m} \vec{N}^{1}_{l,m}(\vec{r})
\end{equation}
with $\vec{M}_{l,m}(\theta, \varphi)$ and $\vec{N}_{l,m}(\theta, \varphi)$ being VSWFs.
VSWFs are defined as
\begin{equation}
\begin{aligned}
\vec{M}_{l,m}^{1,3}(k \vec{r}) &=z_{n}^{1,3}(k r) \vec{m}_{l,m}(\theta, \varphi) \\
\vec{N}_{l,m}^{1,3}(k \vec{r}) &=\sqrt{\frac{n(n+1)}{2}} \frac{z_{n}^{1,3}(k r)}{k r} Y_{l,m}(\theta, \varphi) \vec{e}_{r}+\frac{\frac{\text{d}}{\text{d}kr} [kr z_{n}^{1,3}(k r)]}{k r} \vec{n}_{l,m}(\theta, \varphi)
\end{aligned}
\end{equation}
with $\vec{m}_{l,m}(\theta, \varphi)$ and $\vec{n}_{l,m}(\theta, \varphi)$ being VSH, $Y_{l,m}$ being scalar spherical harmonics as defined by Doicu et al. \cite{Doicu2006}. $z_{n}^{1,3}$ is spherical Bessel functions ($j_n$) for VSWF of the regular type (upper index equal to 1) and spherical Hankel functions ($H_n$) for VSWF of the radiating type (upper index equal to 3).

The incident field expansion coefficients ($a^{mag}_{l,m},a^{el}_{l,m}$) are related to the scattered field expansion coefficients via T-matrix ($T$)
\begin{equation}
    \begin{bmatrix}
    b^{mag} \\ b^{el}
    \end{bmatrix}=
    \begin{bmatrix}
T^{mm} & T^{me} \\
T^{em} & T^{ee}
\end{bmatrix} 
    \begin{bmatrix}
    a^{mag} \\ a^{el}
    \end{bmatrix}
    \label{eq:tmatrix}
\end{equation}

The following orthogonality relations apply to VSHs
\begin{equation}
\begin{aligned}
&\int_{0}^{2 \pi} \int_{0}^{\pi} \vec{m}_{l,m}(\theta, \varphi) \cdot \vec{m}_{l^{\prime},m^{\prime}}^{*}(\theta, \varphi) \sin \theta \mathrm{d} \theta \mathrm{d} \varphi \\
&=\int_{0}^{2 \pi} \int_{0}^{\pi} \vec{n}_{l,m}(\theta, \varphi) \cdot \vec{n}_{l^{\prime},m^{\prime}}^{*}(\theta, \varphi) \sin \theta \mathrm{d} \theta \mathrm{d} \varphi=\pi \delta_{m m^{\prime}} \delta_{l l^{\prime}}
\end{aligned}
\end{equation}
and
\begin{equation}
\int_{0}^{2 \pi} \int_{0}^{\pi} \vec{m}_{l,m}(\theta, \varphi) \cdot \vec{n}_{l^{\prime},m^{\prime}}^{*}(\theta, \varphi) \sin \theta \mathrm{d} \theta \mathrm{d} \varphi=0
\end{equation}
For the scalar spherical harmonics a similar orthogonality relation is 
\begin{equation}
\int_{0}^{2 \pi} \int_{0}^{\pi} Y_{m n}(\theta, \varphi) Y_{m^{\prime} n^{\prime}}(\theta, \varphi) \sin \theta \mathrm{d} \theta \mathrm{d} \varphi=2 \pi \delta_{m,-m^{\prime}} \delta_{n n^{\prime}}
\end{equation}
Using these relations one can show that
\begin{equation}
\int_{0}^{2 \pi} \int_{0}^{\pi} \vec{M}^{3}_{l,m}(\theta, \varphi) \cdot \vec{M}^{3*}_{l,m}(\theta, \varphi) \sin \theta \mathrm{d} \theta \mathrm{d} \varphi = \pi\left(\frac{1}{k^2r^2}|\frac{\text{d}}{\text{d}kr} [kr h(kr)]|^2+l(l+1)\left|\frac{h(kr)}{kr}\right|^2\right)
\label{eq:int1}
\end{equation}
\begin{equation}
\int_{0}^{2 \pi} \int_{0}^{\pi} \vec{N}^{3}_{l,m}(k,\vec{r}) \cdot \vec{N}^{3*}_{l,m}(k,\vec{r}) \sin \theta \mathrm{d} \theta \mathrm{d} \varphi = \pi |h(kr)|^2
\label{eq:int2}
\end{equation}
\begin{multline}
 \int_{0}^{2 \pi} \int_{0}^{\pi} \vec{M}^{3}_{l,m}(k,\vec{r}) \cdot \vec{M}^{1*}_{l,m}(k,\vec{r}) \sin \theta \mathrm{d} \theta \mathrm{d} \varphi =\\
 =\pi \left(\frac{1}{k^2r^2}\frac{\text{d}}{\text{d} kr} [kr j(kr)]^{*} \cdot \frac{\text{d}}{\text{d}kr} [kr h(kr)] +l(l+1)\frac{j(kr)}{kr}^{*}\frac{h(kr)}{kr}\right)
\label{eq:int3}
\end{multline}
\begin{equation}
\int_{0}^{2 \pi} \int_{0}^{\pi} \vec{N}^{3}_{l,m}(k,\vec{r}) \cdot \vec{N}^{1*}_{l,m}(k,\vec{r}) \sin \theta \mathrm{d} \theta \mathrm{d} \varphi = \pi j(kr)^{*} h(kr)
\label{eq:int4}
\end{equation}

\subsection*{Spatial averaging of the optical chirality enhancement}
We use the definition of OCE ($f$) from the main text (eq. 1) and substitute the fields with their decomposition into incident and scattered fields,
\begin{equation}
f^{T}=-Z\text{Im}\left((\vec{E}_{scat}+\vec{E}_{inc})^{*}\cdot(\vec{H}_{scat}+\vec{H}_{inc})\right).
\end{equation}
After evaluating the scalar product, the result is
\begin{equation}
f^{T}=-Z\text{Im}\left(\vec{E}_{scat}^{*}\cdot \vec{H}_{scat}+\vec{E}_{inc}^{*}\cdot \vec{H}_{scat}+\vec{E}_{scat}^{*}\cdot \vec{H}_{inc}+\vec{E}_{inc}^{*}\cdot \vec{H}_{inc}\right).
\end{equation}
We assign a symbol to each term,
\begin{equation}
f^{T}=f^{T}_{scat}+f^{T}_{int,I}+f^{T}_{int,S}+1.
\end{equation}

Orthogonality of VSHs results in canceling out of any terms in which $m \neq m'$ or $l \neq l'$ and that include products of $\vec{M}$ and $\vec{N}$ or their complex conjugates. We assume that the structure is axially symmetric and hence, its T-matrix is diagonal with respect to $m$. The left-handed CPL illumination contains only $m=1$. Thus, from this point we drop index $m$ and assume that it is always $m=1$. Therefore,
\begin{equation}
    f^{T}_{scat,l}=\frac{1}{4\pi}\int_{0}^{2 \pi} \int_{0}^{\pi} {b^{mag}}^{*}_{l} b^{el}_{l} \vec{N}_{l} \cdot  \vec{N}_{l} + {b^{el}}^{*}_{l} b^{mag}_{l} \vec{M}_{l} \cdot  \vec{M}_{l} \sin \theta \mathrm{d} \theta \mathrm{d} \varphi,
\end{equation}
\begin{equation}
    f^{T}_{int,I,,l}=\frac{1}{4\pi}\int_{0}^{2 \pi} \int_{0}^{\pi} {a^{mag}}^{*}_{l} b^{el}_{l} \vec{N}_{l} \cdot  \vec{N}_{l} + {a^{el}}^{*}_{l} b^{mag}_{l} \vec{M}_{l} \cdot  \vec{M}_{l} \sin \theta \mathrm{d} \theta \mathrm{d} \varphi,
\end{equation}
\begin{equation}
    f^{T}_{int,S,,l}=\frac{1}{4\pi}\int_{0}^{2 \pi} \int_{0}^{\pi} {b^{mag}}^{*}_{l} e^{el}_{l} \vec{N}_{l} \cdot  \vec{N}_{l} + {b^{el}}^{*}_{l} a^{mag}_{l} \vec{M}_{l} \cdot  \vec{M}_{l} \sin \theta \mathrm{d} \theta \mathrm{d} \varphi.
\end{equation}

The integrals are evaluated using eqs. \ref{eq:int1}-\ref{eq:int4},
\begin{equation}
    f^{T}_{scat}=\frac{1}{4}\sum_l (b^{mag}_l)^{*}b^{el}_l|h_l(kr)|^2+(b^{el}_l)^{*}b^{mag}_l
    \left(\frac{1}{k^2r^2}\left|\frac{\text{d}}{\text{d}kr} [kr h_l(kr)]\right|^2 + l(l+1)\left|\frac{h_l(kr)}{kr}\right|^2\right),
\end{equation}
\begin{multline}
    f^{T}_{int,I}=\frac{1}{4}\sum_l j_l(kr)^{*} h_l(kr)(a_l^{mag})^{*}b_l^{el} + \\
    \left(\frac{1}{k^2r^2}\frac{\text{d}}{\text{d}kr} [kr j_l(kr)]^{*} \cdot \frac{\text{d}}{\text{d} kr} [kr h_l(kr)] + l(l+1)\frac{j_l(kr)}{kr}^{*}\frac{h_l(kr)}{kr}\right)(a_l^{el})^{*}b_l^{mag},
    \label{eq:finti}
\end{multline}        
\begin{multline}
    f^{T}_{int,S}= \frac{1}{4}\sum_l h_l(kr)^{*}j(kr)(b_l^{mag})^{*}a_l^{el}+ \\
    \left(\frac{1}{k^2r^2}\frac{\text{d}}{\text{d}kr} [kr h_l(kr)]^{*} \cdot \frac{\text{d}}{\text{d}kr} [kr j_l(kr)] +l(l+1)\frac{h_l(kr)^{*}}{kr}\frac{j_l(kr)}{kr}\right)(b_l^{el})^{*}a_l^{mag}.
\end{multline}       
We simplify these equations by inserting the VSWF expansion coefficients for the incident field ($a_l^{mag}=a_l^{el}=\sqrt{2(2l+1)} i^{l-1}$) and use helicity preserving multipoles (eq. 11 in the main text) which leads to the eqs. 13 and 15 in the main text.

\subsection*{Orientation averaging of surface averaged chirality enhancement}
We split the calculation of the orientation averaged chirality enhancement into two steps. First, we find the interference terms ($f^{T}_{int,S}$,$f^{T}_{int,I}$) and then, we find the scattered field contribution ($f^{T}_{scat}$). Each of these terms requires a different orientation averaging procedure.

Calculation of the interference terms requires finding the orientation averaged value of the scattered fields ($\vec{E}_{scat}$,$\vec{H}_{scat}$). We show the procedure for the electric field only as the procedure for the magnetic field is the same. To that end, we insert the T-matrix ansatz (eq. \ref{eq:tmatrix}) into  eq. \ref{eq:escat}
\begin{equation}
    \vec{E}_{scat}=\sum_{l,m,l',m'} (T^{mm}_{l,m,l',m'}a_{l',m'}^{mag}+T^{me}_{l,m,l',m'}a_{l',m'}^{el})\vec{M}_{l,m}+(T^{em}_{l,m,l',m'}a_{l',m'}^{mag}+T^{ee}_{l,m,l',m'}a_{l',m'}^{el})\vec{N}_{l,m}
\end{equation}. Note that here we introduce $m$ again not to miss any possible dependence on it. Next, we perform orientation averaging by replacing the T-matrix with its orientation averaged counterpart defined as 
\begin{equation}
    \langle T^{i,j}_{l,m,l',m'} \rangle =\delta_{m m'}\delta_{l l'}t^{ij}_l
\end{equation}
with
\begin{equation}
    t^{ij}_l=\frac{1}{2l+1}\sum_{m'} T^{ij}_{l,m,l',m'}.
\end{equation}
Finally, we realize that the incident field contains only $m=1$ and perform the summation over $m$, 
\begin{equation}
    \vec{E}_{scat}=\sum_{l,m} \langle b^{mag}_{l,m} \rangle \vec{M}_{l,m} + \langle b^{el}_{l,m} \rangle \vec{N}_{l,m}
\end{equation}
where we introduced $\langle \vec{b} \rangle =\langle T \rangle \vec{a} $. After multiplying by the incident magnetic field, the result is an analogue of eq. \ref{eq:finti} an it undergoes the same transformations leading to eq. 17 in the main text.

Finding $f^T_{scat}$ is more involved as it requires finding the orientation average including ${b^{mag}}^{*}b^{el}$. To that end, we construct a vector of scattered field coefficients $\vec{b}=\begin{bmatrix}
    b^{mag} \\ b^{el}
    \end{bmatrix}$
 and a vector of incident field coefficients $  \vec{a}= \begin{bmatrix}
    a^{mag} \\ a^{el}
    \end{bmatrix}$. Then,
\begin{equation}
   f^T_{scat} = \vec{a}^{\dagger} T^{\dagger} F T \vec{a},
\end{equation}
where 
\begin{equation}
    F = \begin{bmatrix}
0  & |h(kr)|^2 \\
|\frac{\text{d}kr h(kr)}{\text{d}kr}|^2+l(l+1)|\frac{h(kr)}{kr}|^2 &    0
    \end{bmatrix}
\end{equation}
Now, because we are finding the product of scattered fields one cannot immediately perform orientation averaging by replacing $T$ with $\langle T \rangle$. Instead we replace the T-matrix with the rotated T-matrix,
\begin{equation}
    \tilde{T}=R_{-}T R_{+}
\end{equation}
with $R_{-}$ and $R_{+}$ being defined via eq. 1.115 by Doicu et al. \cite{Doicu2006}
After algebraic manipulation, using the fact that $F$ is block diagonal
\begin{equation}
   f^T_{scat} = \vec{a}^{\dagger} R_{-}T^{\dagger} F T R_{+}\vec{a}.
   \end{equation}
We find the orientation average of $W=T^{\dagger} F T$, using the procedure outlined by Doicu et al. \cite{Doicu2006} (see the paragraph leading to eq. 1.123),
\begin{equation}
    \langle (W)^{i,j}_{l,m,l',m'} \rangle =\delta_{m m'}\delta_{l l'}\hat{t}^{ij}_l
\end{equation}
with
\begin{equation}
    \hat{t}^{ij}_l=\frac{1}{2l+1}\sum_{m'} (W)^{ij}_{l,m,l',m'}
\end{equation}
Then,
\begin{equation}
   \langle f^T_{scat} \rangle = \vec{a}^{\dagger} \langle W \rangle \vec{a}.
   \end{equation}

\bibliographystyle{unsrt}
\bibliography{sample}